# Optical Analogue of Quantum Spin Hall Effect in a Two-Dimensional Gyrotropic Photonic Crystal

Cheng He*[1], Xiao-Chen Sun*[1], Xiao-ping Liu[1], Zi-Wen Liu[1], Yulin Chen[2], Ming-Hui Lu†[1], Yan-Feng Chen†[1]

[1]National Laboratory of Solid State Microstructures& Department of Materials Science and Engineering, Nanjing University, Nanjing 210093, China

[2]Clarendon Laboratory, Department of Physics, University of Oxford, Parks Road, Oxford, OX1 3PU, UK

* These authors contributed equally to this work

† Correspondence and request for materials should be addressed to M. H. Lu (luminghui@nju.edu.cn) and Y. F. Chen (yfchen@nju.edu.cn)

**We propose an optical counterpart of the quantum spin Hall (QSH) effect in a two-dimensional photonic crystal composed of a gyrotropic medium exhibiting both gyroelectric and gyromagnetic properties simultaneously. Such QSH effect shows unidirectional polarization-dependent transportation of photonic topological edged states, which is robust against certain disorders and impurities. More importantly, we find that such unique property is not protected by conventional time-reversal symmetry of photons obeying the Bosonic statistics $T_c^2 = 1$ but rather by the same symmetry, $T^2 = -1$, as electron's time-reversal symmetry. Based on the tight-binding approximation approach, we construct an effective Hamiltonian for this photonic structure, which is shown to have a similar form to that of an electronic QSH system. Furthermore, the $Z_2$ invariant of such**

**model is calculated in order to unify its topological non-trivial character. Our finding provides a viable way to exploit the optical topological property, and also can be leveraged to develop a photonic platform to mimic the spin properties of electrons.**

*These authors contributed equally to this work

† M. H. Lu (luminghui@nju.edu.cn) and Y. F. Chen (yfchen@nju.edu.cn)

The topology of electronic systems has recently attracted much attention in condensed matter physics research because of its unique properties including robust unidirectional chiral edge states in quantum Hall system [1, 2] and time-reversal (TR) symmetry protected momentum-dependent spin polarization of the surface/edge states in quantum spin Hall (QSH) system [3-5]. The discovery of QSH system has led to a new state of matter, i.e. topological insulators [6, 7], in which robust non-zero spin current together with zero net electric current is supported by the helical spin gapless edge states resulted from the strong spin-orbit coupling.

Motivated by these novel physics, researchers have begun to explore non-trivial photonic topological states due to the inherent similarity between electrons and photons. The optical analogue of the quantum Hall effect has recently been proposed and demonstrated by the presentence of robust one-way propagation of the TE/TM (electric/magnetic field out of the plane) polarized light in gyroelectric/gyromagnetic two-dimensional photonic crystals (2DPCs) with broken TR symmetry by an external magnetic field [8-11]. Two-dimensional arrays of coupled resonator optical waveguides with TR symmetry have also been suggested recently for achieving topologically protected optical delay lines [12].

Despite all these progress, demonstrating an optical analogue of the QSH effect remains challenging, because unlike electrons, photons obey Bosonic statistics. Consequently, realizing photonic Kramers degenerate states guaranteed by the TR invariance is difficult. A potential and promising method in realizing photonic QSH effect is to utilize a two-fold degenerate band structure along with a TR invariant conjugate pair of states to form ultimately a four-fold degenerate Dirac point. Such method may involve using two coupled states resulted from the coupling between propagation direction and polarization degrees of freedom, in which case each coupled state experiences opposite effective magnetic gauge fields or effective magnetic fields for photons [12-14].

Recently, Khanikaev et al. presented an approach of achieving photonic topological insulator in a triangle photonic crystal composed of magneto-electric coupled metamaterials [15]. Two gapless hybrid states are constructed, constituting Kramers partners of the photonic states, analogous to spin-up and spin-down states in electronic systems. These two topological edged states with opposite unidirectional propagation are connected via TR symmetry, and are robust against disorders or vacancies. Such exploratory study has established a pathway to achieve the photonic QSH effect, and inspired researchers for further investigation using different media and/or structural symmetry. However, it is found that some time-reversal impurities, such as chiral or pure dielectric impurities, would break the backscattering immune property of the above mentioned structure. Thus, the symmetry, which is responsible for the protection of the robustness of photonic topological states, is still unclear and remains to be investigated.

In this letter, we propose an alternative optical model to simulate QSH effect using a square-lattice two-dimensional gyrotropic photonic crystal exhibiting both gyroelectric and gyromagnetic properties simultaneously. We construct a low energy effective Hamiltonian for such system based on the tight-binding approximation (TBA) approach, which is equivalent to that of the electronic QSH system. Further calculations demonstrate the existence of a pair of boundary states responsible for the polarization-dependent transportation of the TE and TM mode. These two polarized waves have definite (but opposite) propagating directions along the photonic crystal boundary, leading to zero net energy flow. By analyzing the symmetry in this system, we find that certain TR chiral impurities would invalidate such QSH states, indicating that the conventional

$T_c^2 = 1$ TR symmetry is not valid to protect the robustness of such QSH states, instead the robustness is rather protected by a new symmetry, which might be the general requirement for photonic crystal QSH systems. Furthermore, the $Z_2$ invariant calculation confirms the non-trivial topological character of the QSH states in our system. Finally, we also discuss the feasibility of constructing such photonic structure with gyrotropic response.

Assuming that there is a gyrotropic material exhibiting unique relative permittivity and permeability in the xy plane, which can be expressed as:

$$\vec{\varepsilon} = \begin{bmatrix} \varepsilon_d & i\varepsilon_f & 0 \\ -i\varepsilon_f & \varepsilon_d & 0 \\ 0 & 0 & \varepsilon_\perp \end{bmatrix}, \vec{\mu} = \begin{bmatrix} \mu_d & i\mu_f & 0 \\ -i\mu_f & \mu_d & 0 \\ 0 & 0 & \mu_\perp \end{bmatrix}. \tag{1}$$

To construct a two-fold degenerate band structure, the parameters are chosen as follows, $\varepsilon_d = \mu_d, \varepsilon_f = -\mu_f, \varepsilon_\perp = \mu_\perp$. Analyzing the edge states of a three-layer sandwiched waveguide, as shown in Fig. 1(a), is straightforward [16, 17]. For simplicity, a pure imaginary refractive index medium (e.g., metal) is used as a waveguide cladding/boundary layer to confine gapless edge states along the waveguide-cladding interface. Since the cladding layer has little influence on the bulk photonic band structures, its optical property can be tailored to facilitate the formation of pure gapless edge states [16]. Other types of boundaries can also be utilized for the same purpose, such as a gyromagnetic photonic crystal-air interface [17, 18].

In detail, the configuration of the supercell for the edge states calculation as shown in Fig. 1(a) contains a periodic gyrotropic photonic crystal bounded by a left/right cladding layer with a refractive index of -i. Our simulation utilizes measured parameters of Yttrium-Iron-Garnet(YIG) crystal under 1600 Gauss external magnetic field at 4.28 GHz RF frequency, i.e., $\mu_d = 14, \mu_f = 12.4, \varepsilon_\perp = 15$ [19]. The rest of the parameters are determined as follows,

$\varepsilon_d = 14, \varepsilon_f = -12.4, \mu_\perp = 15$. A set of vectorial orthogonal eigen equations for TE and TM modes can be simplified to the following set of equations containing only z field components:

$$\hat{H}\begin{bmatrix} E_z \\ H_z \end{bmatrix} = \begin{bmatrix} \mathcal{L}_0 - i\mathcal{L}_1 & 0 \\ 0 & \mathcal{L}_0 + i\mathcal{L}_1 \end{bmatrix}\begin{bmatrix} E_z \\ H_z \end{bmatrix} = 0, \tag{2}$$

where

$$\begin{aligned}
\mathcal{L}_0 &= k_0^2 \varepsilon_\perp + \partial_x \left( \frac{\mu_d}{\mu_d^2 - \mu_f^2} \right) \partial_x + \partial_y \left( \frac{\mu_d}{\mu_d^2 - \mu_f^2} \right) \partial_y \\
&= k_0^2 \varepsilon_\perp + \partial_x \left( \frac{\varepsilon_d}{\varepsilon_d^2 - \varepsilon_f^2} \right) \partial_x + \partial_y \left( \frac{\varepsilon_d}{\varepsilon_d^2 - \varepsilon_f^2} \right) \partial_y \\
\mathcal{L}_1 &= \partial_x \left( \frac{\mu_f}{\mu_d^2 - \mu_f^2} \right) \partial_y - \partial_y \left( \frac{\mu_f}{\mu_d^2 - \mu_f^2} \right) \partial_x \\
&= \partial_x \left( \frac{\varepsilon_f}{\varepsilon_d^2 - \varepsilon_f^2} \right) \partial_y - \partial_y \left( \frac{\varepsilon_f}{\varepsilon_d^2 - \varepsilon_f^2} \right) \partial_x
\end{aligned}.$$

Detailed numerical investigations of the above equations are conducted using commercial FEM software (COMSOL MULTIPHYSICS 3.3).

Fig. 1(b) shows the projected band structures using cladding layer parameters: $\mu_d^c(\varepsilon_d^c) = -1$ and $\mu_\perp^c(\varepsilon_\perp^c) = 1$. The blue line represents the degenerate bulk states of TE and TM mode, and the green and red lines between the second and third energy bands represent two edge states. Figs. 1(c) and 1(d) show the field distribution for point **c** and **d** in Fig. 1(b), respectively. At the frequency corresponding to the point **c**, localized electric (magnetic) field distribution at the left (right) interface along z direction suggests that the left (right) interface only supports TE (TM) mode. Notice that the dispersion curve's slope is always positive in this case, indicating the light travels in a forward direction. However, the properties discussed above for point **c** are completely reversed for point **d**. Therefore, only clockwise (counterclockwise) scattering for TM (TE) wave is permitted along the whole boundary (clockwise and counterclockwise directions are defined by the right hand thumb rule along the z-axis).

To illustrate the polarization-dependent transportation by the gapless edge states, we show in Fig. 2 the field distribution and energy flow direction of a "ring-shaped" structure excited by a circularly polarized source $H_z \exp(-i\omega t) + iE_z \exp(-i\omega t)$. In this figure, the white bars indicate the excitation source with the length of $0.5a$. At a frequency of $0.53442 (2\pi c/a)$, TM (TE) mode has only clockwise (counterclockwise) energy flow as illustrated in Fig. 2(c) (Fig. 2(d)). Therefore, the net energy flow for these two modes cancels out, because the energy flows have identical magnitude but opposite directions as shown in Figs. 2(a) and 2(b). Such transportation behavior is similar to that of the electronic edge states with opposite spin directions in the electronic QSH state.

It should be noticed that the TR operators for electrons and photons are different. In electronic system, the spin is the intrinsic property of electrons and the TR operator is $T = i\tau_y K$ ($\tau_y = \begin{bmatrix} 0 & -i \\ i & 0 \end{bmatrix}$ is Pauli matrix and $K$ is complex conjugation). The only way to flip the spin is to break such symmetry with $T^2 = -1$. Therefore, the electronic quantum spin Hall effect is robust against any non-magnetic (TR symmetry invariant of electrons) disorder or impurity. Meanwhile considering the conventional TR operator of photons $T_c = \tau_z K$ ($\tau_z = \begin{bmatrix} 1 & 0 \\ 0 & -1 \end{bmatrix}$ is Pauli matrix), $T_c \begin{bmatrix} E \\ H \end{bmatrix} = \begin{bmatrix} E^* \\ -H^* \end{bmatrix}$ satisfies Maxwell's equations with time $t$ replaced by $-t$. However, any two orthogonal photonic modes can be converted to each other by using some medium obeying such conventional time reversal symmetry ($T_c^2 = 1$). For example, a TE mode can evolve into a TM mode when propagating in chiral medium, and vice versa. Hence the robustness of QSH states can be checked against a chiral impurity with $D = \ddot{\varepsilon}E + \ddot{\xi}H$ and $B = \ddot{\mu}E + \ddot{\zeta}H$, where $\ddot{\varepsilon} = I, \ddot{\mu} = I, \ddot{\xi} = -\ddot{\zeta} = diag\{0, 0, i\}$. In this case, the eigen equations of such chiral

medium can be described as

$$\hat{H}_c \begin{bmatrix} E_z \\ H_z \end{bmatrix} = \begin{bmatrix} k_0^2 + \partial_x^2 + \partial_y^2 & ik_0^2 \\ -ik_0^2 & k_0^2 + \partial_x^2 + \partial_y^2 \end{bmatrix} \begin{bmatrix} E_z \\ H_z \end{bmatrix} = 0. \quad (3)$$

As shown in Fig. 3(a), a chiral impurity with its radius $r = 0.11a$ is placed at the center of the waveguide. With the TE polarized source (represent by black star), the backscattered and propagating TM mode can be excited at the chiral impurity site. Although the TE and TM modes still propagate in opposite directions, the robustness is broken. Fig. 3(b) shows the corresponding band structure, where the Kramers degeneracy disappears, and the edge modes, which are gapless previously, now open a gap. Although the Hamiltonian of chiral medium in Eq. (3) satisfies $T_c \hat{H}_c T_c^{-1} = \hat{H}_c$, the field distribution and band structure in Fig. 3 indicate that the robustness of the net polarization dependent transportation of optical edge states is not protected by the conventional TR symmetry of photons ($T_c^2 = 1$), thus such robustness is broken in this case.

As a result, a new symmetry operator must be defined to describe the symmetry protection mechanism in our system. Considering the degeneracy-symmetry in the system, we find a new operator $T = i\tau_y K$ ($\tau_y = \begin{bmatrix} 0 & -i \\ i & 0 \end{bmatrix}$ is Pauli matrix), which is the same as the TR operator of electrons. The Hamiltonian in Eq. (2) satisfies $T\hat{H}T^{-1} = \hat{H}$, which means the robust property is protected by this new symmetry. And, the chiral impurity mentioned in Eq. (3) does not satisfy such new symmetry, i.e. $T\hat{H}_c T^{-1} \neq \hat{H}_c$, which is consistent with the simulation results in Fig. 3. Therefore, the optical quantum spin Hall model present here is robust against any $T$-symmetric invariant disorder or impurity (more details are in part E of [20]). Moreover, this new symmetry operator satisfies $T^2 = (i\tau_y K)^2 = -1$, which is the same as TR operator of electronic quantum spin Hall effect.

In order to construct an analytical model via the low-energy effective Hamiltonian, TBA approach is used to analyze our system where the TE and TM modes can be treated independently. For the TE mode there are three types of eigen-modes, $|s^{TE}\rangle$, $|p_x^{TE}\rangle$ and $|p_y^{TE}\rangle$, equivalent to an electric monopole and two electric dipole modes in an electronic system. The mode hopping behavior among these eigen-modes is analyzed with TBA method and summarized as follows: The interaction between $|p_x^{TE}\rangle$ and $|p_y^{TE}\rangle$ only contains the on-site coupling, while the interaction between other pairs contains both the on-site coupling and the nearest-neighboring coupling. Herein, we define new $p$ states as the linear superposition of $|p_x^{TE}\rangle$ and $|p_y^{TE}\rangle$ with $|p_\pm^{TE}\rangle = \mp \frac{1}{\sqrt{2}}(|p_x^{TE}\rangle \pm i|p_y^{TE}\rangle)$. By applying the perturbation theory and expanding the k dependent Hamiltonian to the first order, we obtain:

$$H_{TE} = \epsilon_0 + m_0 \tau_z + \hbar v_f (k_y \tau_x + k_x \tau_y) \to H_0 \tag{4}$$

where $\tau_x$, $\tau_y$ and $\tau_z$ are Pauli matrices, $\epsilon_0$ is the energy origin point, $m_0$ is mass term and $v_f$ represents the phase velocity near the Dirac point (see the part A and B of [20] for details). Clearly the Hamiltonian here has a Dirac form.

The Hamiltonian of one polarization state can therefore be expressed in terms of the other polarization state via symmetry operation on all eigen-modes. The total Hamiltonian can then be written as

$$H = \begin{bmatrix} H_0(\vec{k}) & \\ & H_0^*(-\vec{k}) \end{bmatrix}, \tag{5}$$

where $H_0$ is defined in Eq. (4). This total Hamiltonian representation is exactly the same as the Hamiltonian in BHZ model of 2D electronic QSH effect [21], which remains invariant under TR operation. By solving the eigen-equations Eq. (5) (see the part C of [20] for details), the dispersion

of the edge states for our system can be determined as:

$$E = \sigma \hbar v_f k_y, \sigma = \pm 1. \tag{6}$$

Fig. 4 shows this edge-state dispersion (solid line), which agrees well with the simulation data (circles). Notice that in the calculation $\hbar v_f = 0.1153$ was chosen for the edge states.

Eq. (5) can also be used to calculate $Z_2$ invariant. Due to the inversion symmetry of the model, the $Z_2$ invariant can be determined by the quantities $\delta_i = \prod_{m=1}^{N} \xi_{2m}(\Gamma_i)$ [22]. Here, $\xi_{2m}(\Gamma_i)$ is the parity eigenvalue of the 2m$^{th}$ occupied energy band at four $T$ invariant (which has the same effect as time-reversal invariant) momenta $\Gamma_i$ in the Brillouin zone. The $Z_2$ invariant $\nu = 0, 1$, which distinguishes the quantum spin-Hall phase, is governed by the product of all the $\delta_i$: $(-1)^\nu = \prod_i \delta_i$. As for the $4 \times 4$ matrix discussed Eq. (5), the quantities equation can further be simplified as $\delta_i = \xi(\Gamma_i)$, in which $\xi(\Gamma_i)$ is the parity eigenvalue of lower energy band at $\Gamma_i$. Because of $T$ invariant, the eigen states have degeneracy at $\Gamma_i$. The Bloch field distributions of four high symmetric points are shown in Fig. 5. The arrows show the direction of energy flow. We can find that the Bloch field distribution at (0, 0) point is even parity with parity eigenvalue 1, while the Bloch field distributions at other three points are odd parity with parity eigenvalues -1. Therefore, $(-1)^\nu = -1$ and the $Z_2$ invariant is $\nu = 1$, which imply the non-trivial topological states.

The spin of electron in the QSH effect in electronic systems is described as Dirac fields with $T_c^2 = -1$. According to Kramers theorem, the single-particle eigen-state of the Hamiltonian must have a degenerate partner. When one-way propagating electrons are back scattered by nonmagnetic impurity, the two possible backscattering paths always interfere destructively, leading to perfect transmission [23]. Similarly, in this photonic system, the $T$ operator with

$T^2 = -1$ means that rigorous Kramers double degeneracy can exist for photons, implying the transportation robustness of the net polarization carried by optical edge states can be protected by such symmetry. Furthermore, the robustness of polarization-dependent edge states against the disorder and impurity in photonic systems can exist. Thus, the edge states are still robust in our system and they exhibit the same back scattering immune property as electrons in the QSH model. Detailed simulation results for this part are shown in the part E of [20].

Homogenous gyrotropic material used in our photonic system may not exist naturally. However, there are several ways to obtain an effective gyrotropic response. I) An effective medium combining two sub-wavelength sized naturally-occurring gyroelectric and gyromagnetic materials [10] can be formed to mimic gyrotropic response. II) There exist metamaterials with unique electromagnetic responses, including magnetic resonance, electronic resonance, optical active chirality, and magneto-electric coupling in structures such as split-ring resonators [15, 24], metallic-dielectric core-shell particles [25] and metallic helices [26]. Gyrotropic response in an artificial metamaterial medium might also be engineered in desired frequency ranges. III) There are designs that can generate an effective magnetic field by controlling the optical phase dynamic in photonic crystals and meta-materials [14]. In addition, in some artificial multiferroic materials, magnetic field could be applied to manipulate ferroelectric polarization, and vice versa. These effects could be leveraged to create a material system with the coexistence of off-diagonal terms in $\varepsilon$ and $\mu$ matrices [27].

To engineer the cladding layer's electromagnetic response, some designs can be utilized including periodic arrays of interspaced conducting nonmagnetic split ring resonators and continuous wires, which exhibit negative values for the z (out of plane) component of $\varepsilon$ and $\mu$ tensor [28]. Notice that working with gigahertz frequency offers great feasibility in engineering the required gyrotropic response and frequency scaling into optical window remains challenging due to the complexity in nanofabrication

In summary, we have numerically realized an optical QSH in a 2DPC system exhibiting both gyroelectric and gyromagnetic properties simultaneously. The backscattering immune optical polarization-dependent transportation can be achieved in a pair of conjugate gapless edged states. More importantly, we demonstrate clearly that the robustness of such QSH is protected by the symmetry operator $T = i\tau_y K$ (the same as the TR operator of electrons), which might be the general symmetry requirement in order to use photonic polarization to mimic electronic spin property. The low energy effective Hamiltonian is retrieved by the TBA method, the versatility of which is clearly demonstrated in the construction of the topological edged states with different basis states. Moreover, the $Z_2$ invariant in our model is discussed to classify the non-triviality of the topological states for our system. Our work provides additional physical insights for topological-phase related research and establishes a new way to study the topological phenomenon in $T^2 = -1$ photonic crystal systems. In addition, our work may find great potential applications such as polarization splitter for entangled photons [29], optical isolation and polarization-dependent transportation.

The work was jointly supported by the National Basic Research Program of China (Grant No. 2012CB921503 and No. 2013CB632702) and the National Nature Science Foundation of China (Grant No. 1134006). We also acknowledge the support from Academic Program Development of Jiangsu Higher Education (PAPD) and China Postdoctoral Science Foundation (Grant No.

2012M511249 and No. 2013T60521). Y.L. Chen acknowledges support from a DARPA MESO project (No. 187 N66001-11-1-4105).

Figures caption:

Fig. 1 (Color online). (a) Supercell configuration containing periodical gyrotropic photonic crystal bounded by a left/right cladding layer with a refractive index of -i. (b) The projected band structures of circularly polarized wave (left panel). Gapless edge modes are denoted by green and red lines. Cladding layer parameters are $\mu_d^c(\varepsilon_d^c) = -1$ and $\mu_\perp^c(\varepsilon_\perp^c) = 1$. The right panel is a zoom-in view of the rectangular frequency area. (c) and (d) correspond to the field distribution of points c and d in Fig 1.(b), respectively. Notice that both electric and magnetic fields are plotted.

Fig. 2 (Color online). (a) A schematic of the total energy flow in a ring-shaped PC. (b) The net zero energy flow for TE and TM modes combined. (c) The clockwise propagating energy flow of TM mode. (d) The anti-clockwise propagating energy flow of TE mode. The white bar represents the excitation source. Red and blue arrows represent the energy flow of TM and TE modes, respectively.

Fig. 3 (Color online). (a) Field distribution with a chiral impurity. (b) Bandstructure of square-lattice gyrotropic photonic crystal with optical chirality.

Fig. 4 (Color online). Dispersion relation for edge states. Solid line and open circles represent the theoretical result according to Eq. (6) and simulation data, respectively.

Fig. 5 (Color online). The Bloch field distribution of four high symmetric points in the first Brillouin zone. The arrows indicate the direction of energy flow.

Reference:

[1] Y. Hatsugai, Phys. Rev. Lett. **71**, 3697 (1993).

[2] D. J. Thouless, M. Kohmoto, M. P. Nightingale, and M. Dennijs, Phys. Rev. Lett. **49**, 405 (1982).

[3] C. L. Kane and E. J. Mele, Phys. Rev. Lett. **95**, 146802 (2005).

[4] J. Wunderlich, B. Kaestner, J. Sinova, and T. Jungwirth, Phys. Rev. Lett. **94**, 047204 (2005).

[5] D. Hsieh, D. Qian, L. Wray, Y. Xia, Y. S. Hor, R. J. Cava, and M. Z. Hasan, Nature **452**, 970 (2008).

[6] Y. Xia, D. Qian, D. Hsieh, L. Wray, A. Pal, H. Lin, A. Bansil, D. Grauer, Y. S. Hor, R. J. Cava, and M. Z. Hasan, Nature Phys. **5**, 398 (2009).

[7] H. J. Zhang, C. X. Liu, X. L. Qi, X. Dai, Z. Fang, and S. C. Zhang, Nature Phys. **5**, 438 (2009).

[8] S. Raghu and F. D. M. Haldane, Phys. Rev. A **78**, 033834 (2008).

[9] Z. Wang, Y. D. Chong, J. D. Joannopoulos, and M. Soljacic, Phys. Rev. Lett. **100**, 013905 (2008).

[10] Z. Wang, Y. Chong, J. D. Joannopoulos, and M. Soljacic, Nature **461**, 772 (2009).

[11] F. D. M. Haldane and S. Raghu, Phys. Rev. Lett. **100**, 013904 (2008).

[12] M. Hafezi, E. A. Demler, M. D. Lukin, and J. M. Taylor, Nature Phys. **7**, 907 (2011).

[13] R. O. Umucalilar and I. Carusotto, Phys. Rev. A **84**, 043804 (2011).

[14] K. Fang, Z. Yu, and S. Fan, Nature Photon. **6**, 782 (2012).

[15] A. B. Khanikaev, S. Hossein Mousavi, W.-K. Tse, M. Kargarian, A. H. Macdonald, and G. Shvets, Nature Mat. **12**, 233 (2013).

[16] C. He, M.-H. Lu, W.-W. Wan, X.-F. Li, and Y.-F. Chen, Solid State Communications **150**, 1976 (2010).

[17] X. Ao, Z. Lin, and C. T. Chan, Phys. Rev. B **80**, 033105 (2009).

[18] Y. Poo, R.-x. Wu, Z. Lin, Y. Yang, and C. T. Chan, Phys. Rev. Lett. **106**, 093903 (2011).

[19] D. M. Pozar, Wiley, 1998, p. 2nd edn.

[20] Supplementary,

[21] B.Bernevig, T.Hughes, and S.-C.Zhang, arXiv:cond-mat **0611399v1**,

[22] L. Fu and C. L. Kane, Phys. Rev. B **76**, 045302 (2007).

[23] X.-L. Qi and S.-C. Zhang, Physics Today **63**, 33 (2010).

[24] R. A. Shelby, D. R. Smith, and S. Schultz, Science **292**, 77 (2001).

[25] W. Liu, A. E. Miroshnichenko, D. N. Neshev, and Y. S. Kivshar, Acs Nano **6**, 5489 (2012).

[26] J. K. Gansel, M. Thiel, M. S. Rill, M. Decker, K. Bade, V. Saile, G. von Freymann, S. Linden, and M. Wegener, Science **325**, 1513 (2009).

[27] S.-W. Cheong and M. Mostovoy, Nature Mat. **6**, 13 (2007).

[28] D. R. Smith, W. J. Padilla, D. C. Vier, S. C. Nemat-Nasser, and S. Schultz, Phys. Rev. Lett. **84**, 4184 (2000).

[29] W. Chen, R. Shen, L. Sheng, B. G. Wang, and D. Y. Xing, Phys. Rev. Lett. **109**, 036802 (2012).

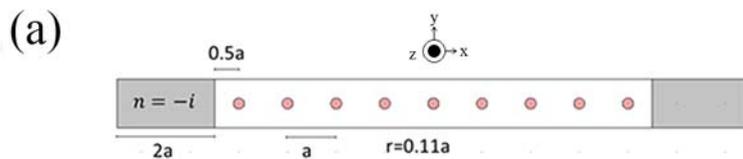

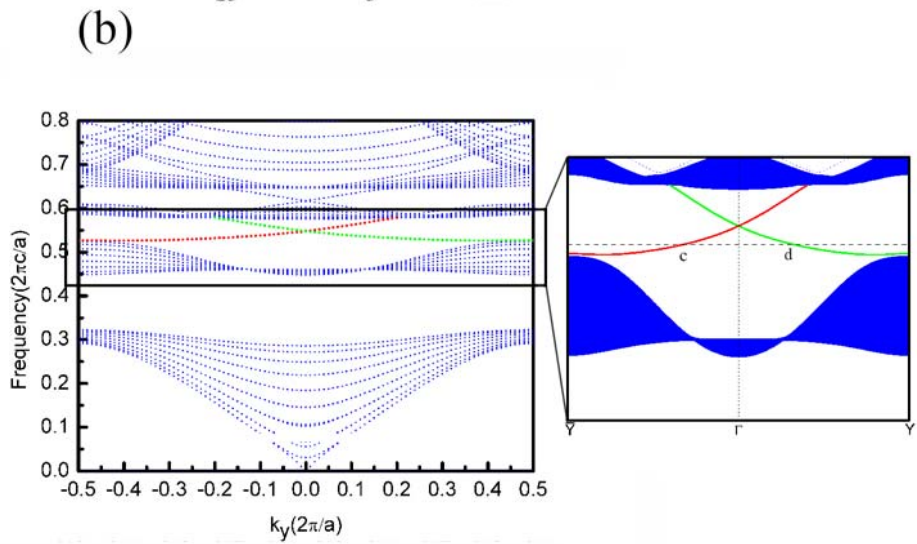

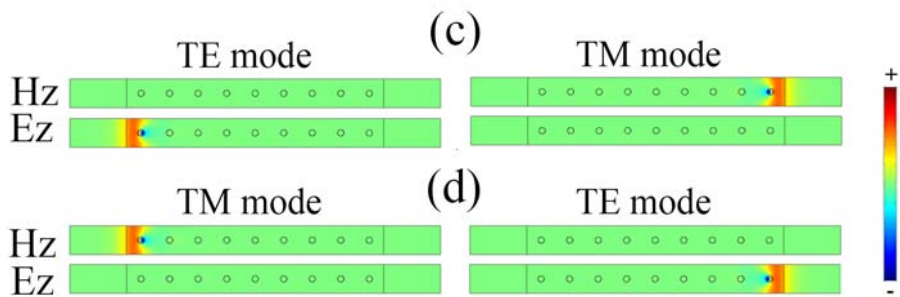

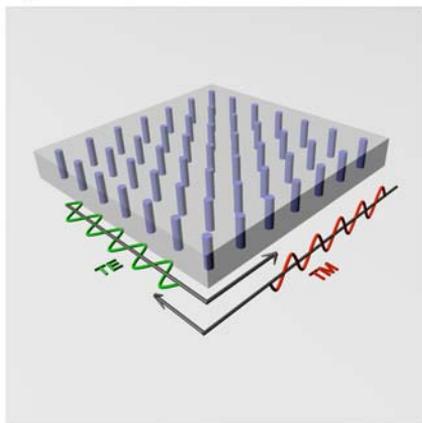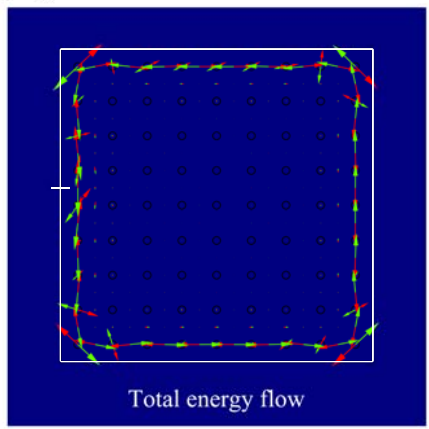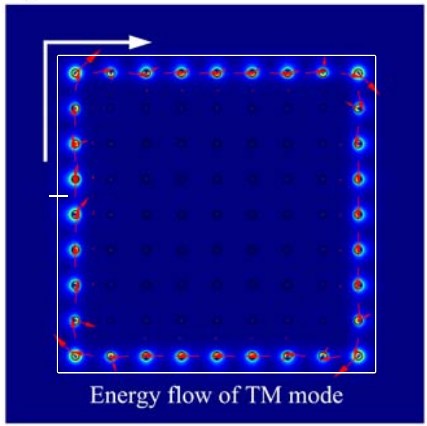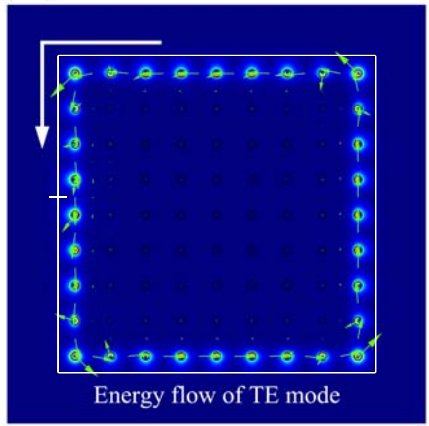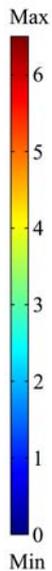

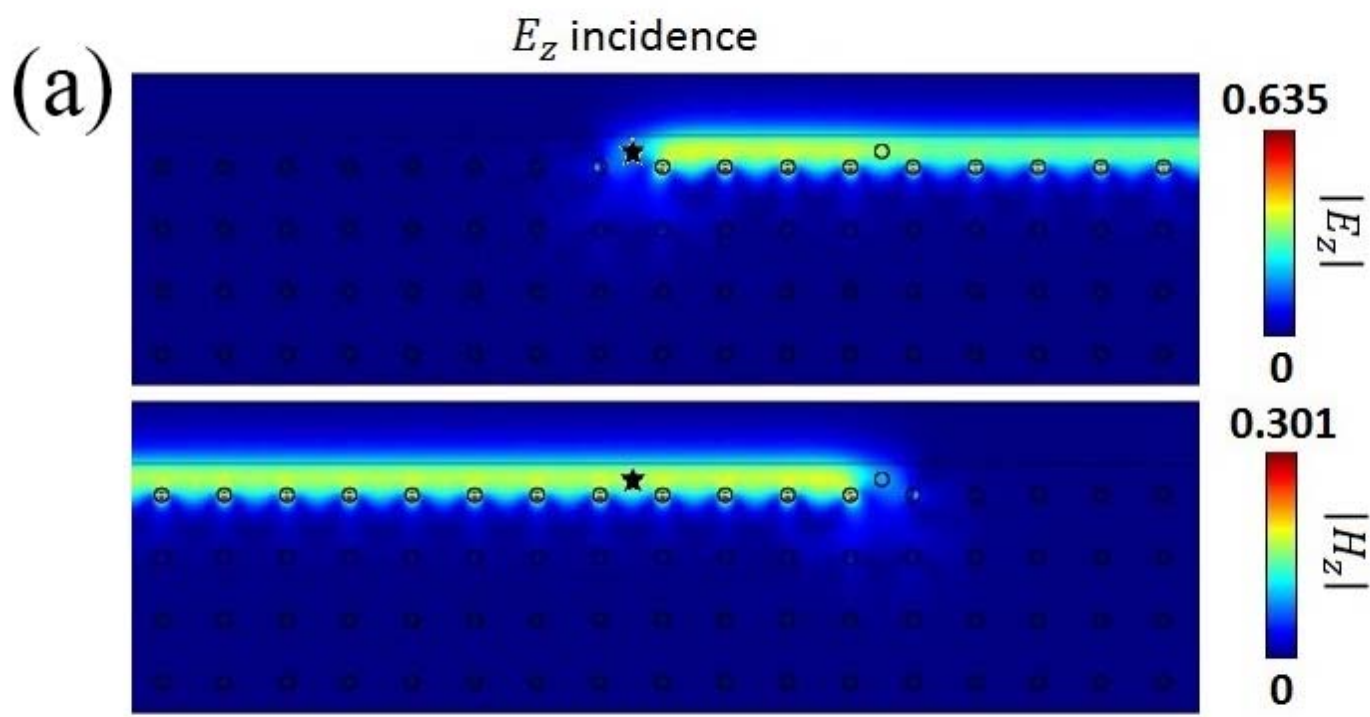

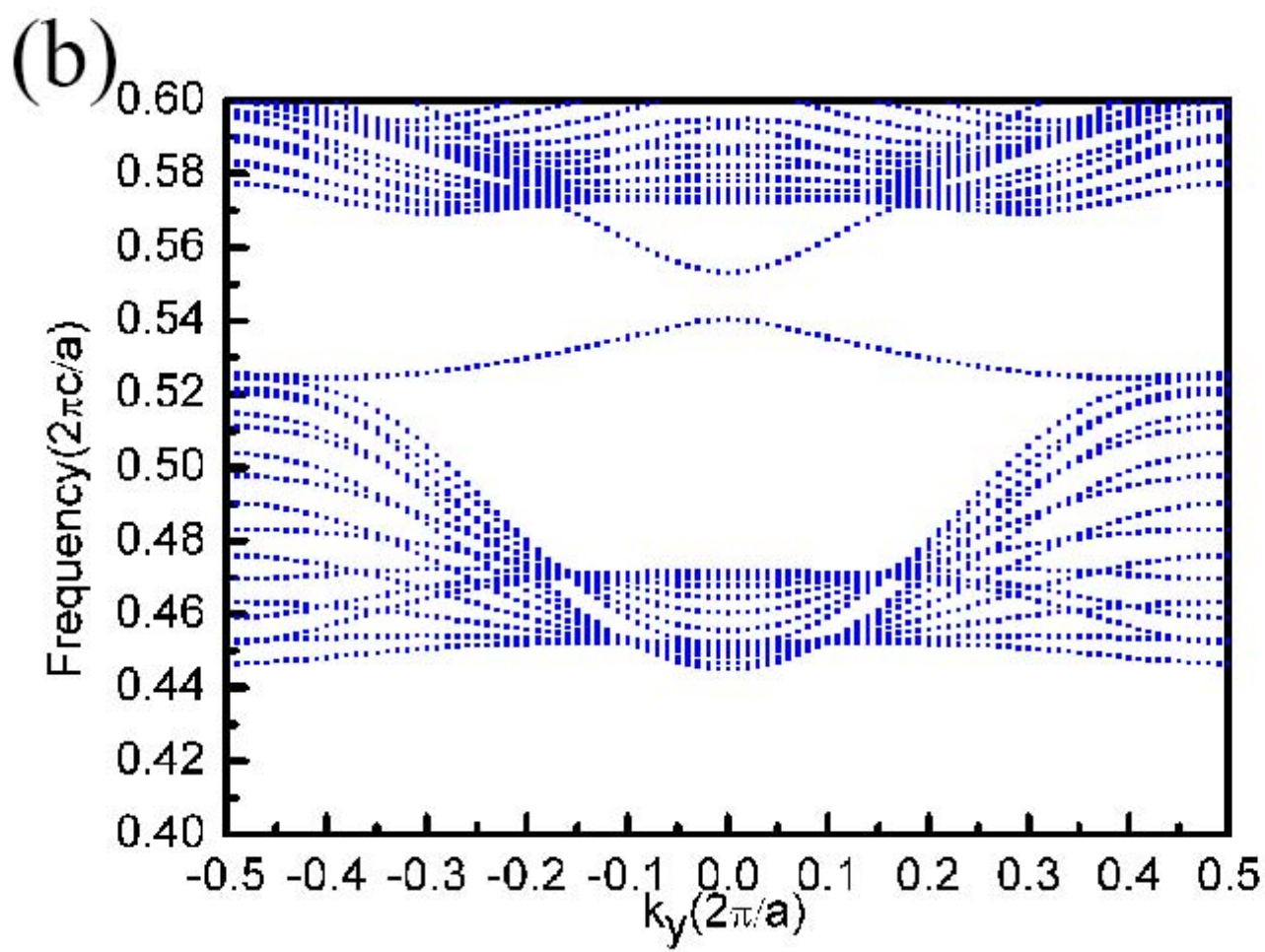

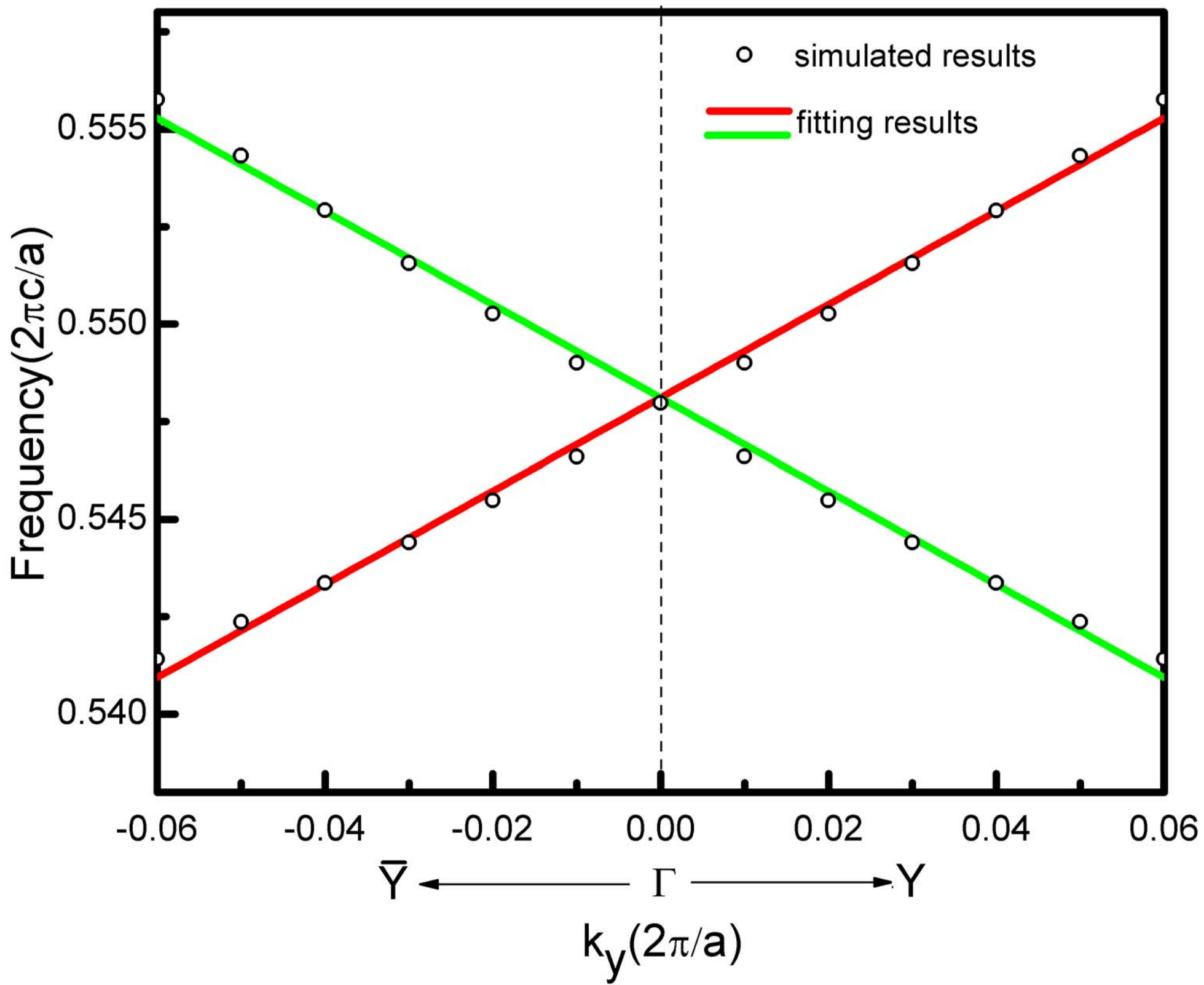

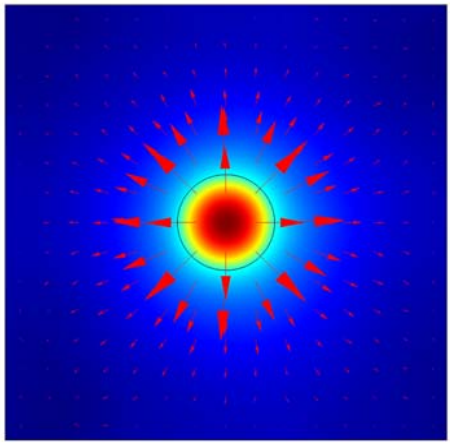

$(0, 0), \delta_{00} = 1$

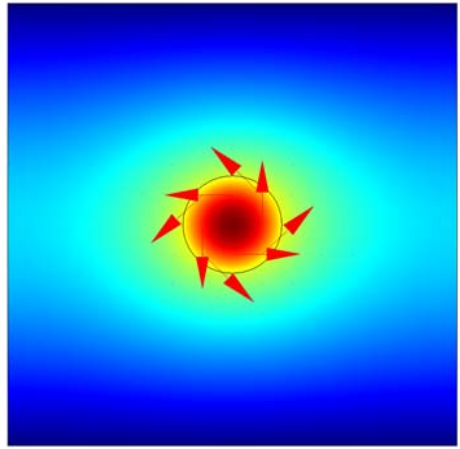

$(0, \pi), \delta_{0\pi} = -1$

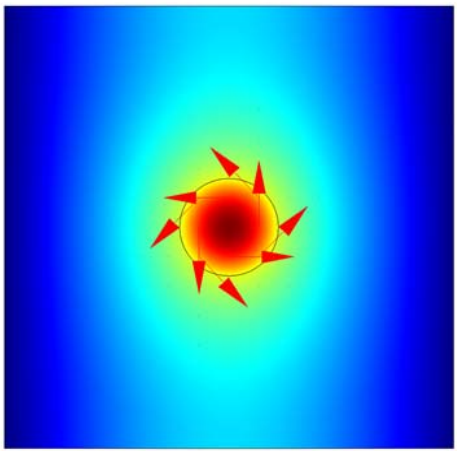

$(\pi, 0), \delta_{\pi 0} = -1$

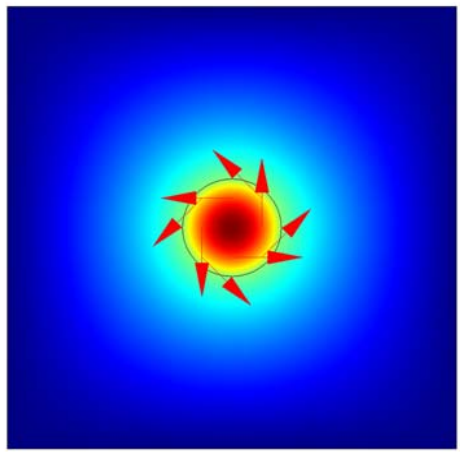

$(\pi, \pi), \delta_{\pi\pi} = -1$